\documentclass[12pt]{iopart}
\usepackage{iopams}
\usepackage{graphicx}

\def\be{\begin{equation}}
\def\ee{\end{equation}}
\def\bea{\begin{eqnarray}}
\def\eea{\end{eqnarray}}
\def\d{\mbox{d}}
\let\phi=\varphi

\def\bh{black-hole }

\let\phi=\varphi
\let\rho=\varrho

\begin{document}

\comment[On non-monotonic orbital velocity profiles around K(a)dS black holes]{On non-monotonic orbital velocity profiles around rapidly rotating Kerr--(anti-)de Sitter black holes: a~comment to the recently published results}

\author{Petr Slan\'{y} and Zden\v{e}k Stuchl\'{\i}k}

\address{Institute of Physics, 
         Faculty of Philosophy and Science, 
         Silesian university in Opava, 
         Bezru\v{c}ovo n\'{a}m. 13, CZ-746 01 Opava, Czech Republic}
           
\ead{Petr.Slany@fpf.slu.cz, Zdenek.Stuchlik@fpf.slu.cz}

\begin{abstract}
Critical discussion of the recently published results [M\"{u}ller~A and Aschenbach~B 2007 \textit{Class. Quantum Grav.} \textbf{24} 2637, \texttt{arXiv:0704.3963}] on the non-monotonic orbital velocity profiles of the Keplerian motion of test particles and $\ell=\mbox{const}$ motion of test perfect fluid around K(a)dS black holes is given, and the discrepancies concerned the existence of the non-monotonicity in dependence of the spacetime parameters are corrected. Moreover a new non-monotonic behaviour of the Keplerian orbital velocity in the Kerr--anti-de Sitter spacetimes is highlighted.
\end{abstract}


\submitto{\CQG}

\maketitle

\section{Introduction}
It has been shown in \cite{Asch:2004:ASTRA:} that the LNRF-related orbital velocity of test particles on Keplerian orbits around the Kerr black hole exhibits a non-monotonic radial behaviour, if the spin of the black hole is $a>0.9953$ (in the dimensionless formulation $c=G=M=1$; $M$ is the \bh mass). Further, it was shown in \cite{Stu-Sla-Tor-Abr:2005:PHYSR4:} that the analogical non-monotonic behaviour can be found also for a non-geodesic circular motion of a~test perfect fluid orbiting with uniformly distributed specific angular momentum\footnote{$\ell(r,\,\theta)=-U_{\phi}/U_{t}$, where $U_{t},\ U_{\phi}$ are covariant components of the 4-velocity field in the fluid connected with the stationarity and axial symmetry of the spacetime.} $\ell(r,\,\theta)=\mbox{const}$ around the Kerr black hole with spin $a>0.99964$.

Similar analysis of the LNRF-related orbital velocity profiles corresponding to the geodesic (Keplerian) and non-geodesic ($\ell=\mbox{const}$) motion of test particles around Kerr--(anti-)\-de~Sitter black holes was done by M\"{u}ller \& Aschenbach in \cite{Mul-Asch:2007:CLAQG:}, stating that an interplay between the cosmological constant $\Lambda$ and the \bh spin $a$ ``seems to control the occurence and modulation depth of the `minimum-maximum' structure''. In the case of Keplerian motion they present the linear fit between the cosmological constant and the critical spin, for which the ``humpy'' behaviour is getting to exist, pointing out that in the KdS black-hole spacetimes the linear relation breaks down for $\Lambda >0.011$ because of $a>1$. Moreover, for the KadS ($\Lambda<0$) black-hole spacetimes they give also the lower limit(?) on the black-hole spin $a>0.991$, allowing existence of the ``minimum-maximum'' structure in the orbital velocity profiles. In the case of $\ell=\mbox{const}$ motion they conclude that the ``minimum-maximum'' structure appears only for $-0.001\lesssim\Lambda\lesssim 0.00017$. In both cases they show that the velocity difference between the emerged local maximum and local minimum of the orbital velocity profile grows with decreasing cosmological constant (from positive to negative values).

In this comment we correct the results of M\"{u}ller \& Aschenbach \cite{Mul-Asch:2007:CLAQG:}. In the case of Keplerian motion we give the corrected relation for the LNRF-related orbital velocity, using the precise K(a)dS-form of the Keplerian angular velocity instead of the Kerr-form used in \cite{Mul-Asch:2007:CLAQG:}, and correct the linear relation between the cosmological constant and the critical spin. Moreover we point out that in KadS spacetimes the Keplerian orbital velocity always reveals the non-monotonic behaviour (although of different quality) independently of the spin $a$. In the case of $\ell=\mbox{const}$ motion we are more careful about the dependence of the ``minimum-maximum'' effect on the value of $\ell$. We obtained results which may contradict those of M\"{u}ller \& Aschenbach \cite{Mul-Asch:2007:CLAQG:}. A clear comparison cannot be done since the $\ell=\mbox{const}$ analysis is not explicitly described in \cite{Mul-Asch:2007:CLAQG:}.

\section{Orbital velocity profiles in K(a)dS black-hole spacetimes}
Kerr--(anti-)de Sitter spacetime, characterized by the central-mass parameter $M$, rotational parameter (black-hole spin) $a>0$ and cosmological constant $\Lambda$, is described by the line element\footnote{The Boyer-Lindquist coordinates ($t,\,r,\,\theta,\,\phi$) and geometrical units ($c=G=1$) are used.}
\bea                                                                          \label{e1}
\d s^2 &=& -\left (\frac{\Delta_{r}\Delta_{\theta}\rho^2}{I^{2}A}\right )\d t^2 +
       \left (\frac{A\sin^{2}\theta}{I^{2}\rho^2}\right )(\d\phi-\omega\d t)^2 \\ \nonumber
       & & + \left (\frac{\rho^2}{\Delta_{r}}\right )\d r^2 + 
       \left (\frac{\varrho^{2}}{\Delta_{\theta}}\right )\d\theta^2,      
\eea
where
\be                                                                           \label{e2}
\Delta_r = r^{2}-2Mr+a^{2}-\frac{1}{3}\Lambda r^2 \left(r^2+a^2 \right),\ 
\Delta_{\theta} = 1+\frac{1}{3}\Lambda a^2 \cos^2 \theta,
\ee
\be                                                                           \label{e3}
\rho^2 = r^2 + a^2 \cos^2 \theta,\quad
A = (r^{2}+a^{2})^{2}\Delta_{\theta}-a^{2}\Delta_{r}\sin^2 \theta,
\ee
\be                                                                           \label{e4}
I = 1+ \frac{1}{3}\Lambda a^2, \quad
\omega = \frac{a}{A}\left [(r^{2}+a^{2})\Delta_{\theta}-\Delta_{r}\right ].
\ee

The orbital velocity of matter is given by appropriate projections of its 4-velocity \textbf{U} onto the tetrad of a locally non-rotating frame (LNRF) \cite{Bar-Pre-Teu:1972:ASTRJ2:,Stu-Sla:2004:PHYSR4:} 
\bea
\mathbf{e}^{(t)}\equiv\left (\frac{\Delta_{r}\Delta_{\theta}\varrho^{2}}{I^{2}A}\right )^{1/2}{\rm d}t,                                                              \label{e5} \\
\mathbf{e}^{(\phi)}\equiv\left (\frac{A\sin^{2}\theta}{I^{2}\varrho^{2}}\right )^{1/2}({\rm d}\phi-\omega {\rm d}t),                                                      \label{e6} \\
\mathbf{e}^{(r)}\equiv\left (\frac{\varrho^{2}}{\Delta_{r}}\right )^{1/2}{\rm d}r, 
                                                                              \label{e7} \\
\mathbf{e}^{(\theta)}\equiv\left (\frac{\varrho^{2}}{\Delta_{\theta}}\right )^{1/2}{\rm d}\theta,                                                                     \label{e8} 
\eea
and corresponds to the locally measured azimuthal component of 3-velocity in the LNRF
\be                                                                           \label{e9}
{\cal V}^{(\phi)}=\frac{U^{\mu}\mathrm{e}^{(\phi)}_{\mu}}{U^{\nu} \mathrm{e}^{(t)}_{\nu}} 
=\frac{A\sin\theta}{\rho^{2}\sqrt{\Delta_{r}\Delta_{\theta}}}(\Omega-\omega),
\ee
where $\omega = -g_{t\phi}/g_{\phi\phi}$, given by (\ref{e4}), is the angular velocity of the LNRF, and 
\be                                                                           \label{e10}
\Omega = \frac{U^{\phi}}{U^{t}} = -\frac{\ell g_{tt}+g_{t\phi}}{\ell g_{t\phi}+g_{\phi\phi}}
\ee
is the angular velocity of a matter orbiting with the specific angular momentum $\ell=-U_{\phi}/U_{t}$. 

Instead of $\Lambda$ we introduce the ``cosmological parameter''
\be                                                                           \label{e11}
     y = \frac{1}{3} \Lambda M^2,
\ee
and reformulate all relations into the completely dimensionless form by putting $M=1$ hereafter.

\subsection{Keplerian distribution of the specific angular momentum}

\begin{figure}
  \begin{minipage}{.49 \linewidth}
    \centering
    \vspace{6pt}
    \includegraphics[width=1 \hsize]{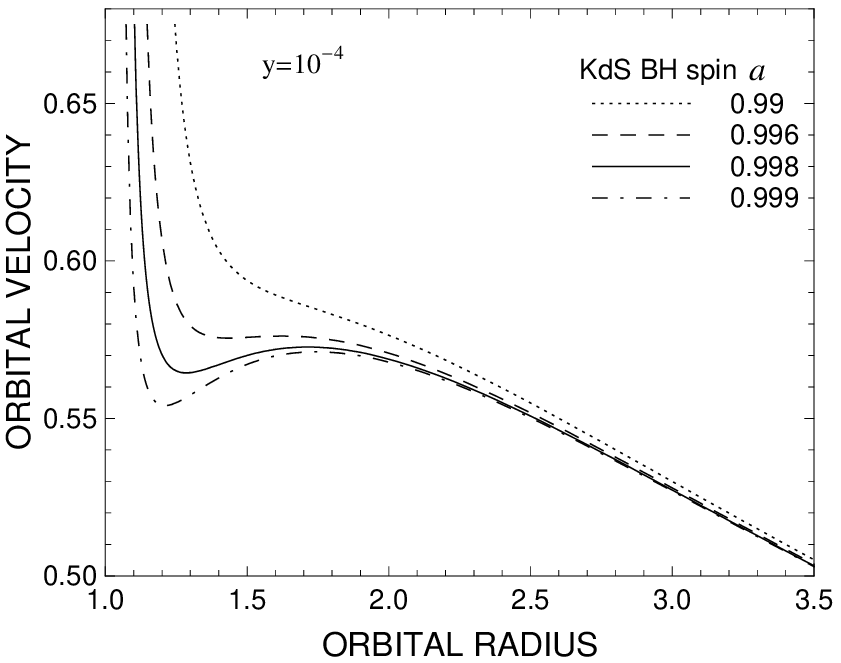}
  \end{minipage}
  \hfill
  \begin{minipage}{.49 \linewidth}
    \centering
    \includegraphics[width=1 \hsize]{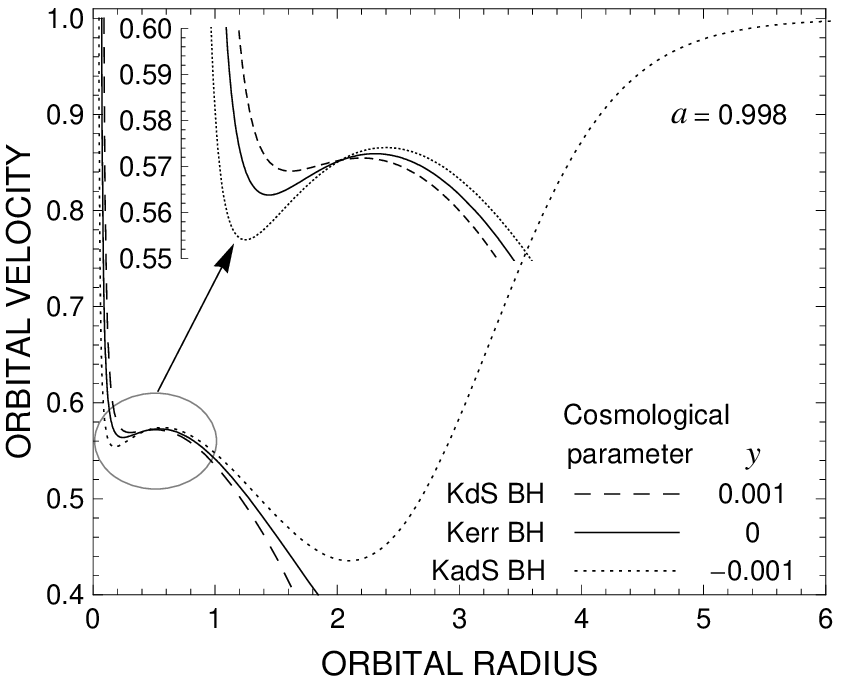}
  \end{minipage}
	\caption{Keplerian orbital velocity profiles in the LNRF. Left panel shows the change of the profile with increasing black-hole spin $a$ and the fixed cosmological parameter $y$ (cosmological constant $\Lambda$). Right panel shows the change of the profile with decreasing cosmological parameter $y$ and the fixed value of the black-hole spin $a$, giving the humpy orbital velocity profiles in the selected K(a)dS spacetimes. The outer minimum of the profile in the Kerr--\emph{anti}-de Sitter background is caused by the attractive cosmological constant ($\Lambda<0$) and exists independently of the spin $a$.}
	\label{f1}
\end{figure}

Motion of test particles following circular orbits in the equatorial plane ($\theta=\pi/2$) is described by the Keplerian distribution of the specific angular momentum, which in the K(a)dS backgrounds takes the form \cite{Stu-Sla:2004:PHYSR4:,Sla-Stu:2005:CLAQG:}
\be                                                                           \label{e12}
\ell_{\rm K\pm}(r;\,a,\,y) = \pm\frac{(r^2+a^2)\sqrt{1-yr^3}\mp
           a\sqrt{r}[2+r(r^2+a^2)y]}{r\sqrt{r}[1-(r^2+a^2)y]-2\sqrt{r}\pm a\sqrt{1-yr^3}};
\ee
$\pm$ refers to two distinct families of orbits in the K(a)dS spacetimes. From the LNRF point of view the minus-family represents retrograde orbits only, while the plus-family can be both direct or retrograde; see the paper of Stuchl\'{\i}k \& Slan\'{y} (2004) for more details.

Corresponding Keplerian angular velocity is given by the relation \cite{Sla-Stu:2005:CLAQG:}
\be                                                                           \label{e13}
\Omega_{\rm K\pm}(r;\,a,\,y) = \frac{1}{a \pm\sqrt{r^3/(1-yr^3)}},
\ee
and the Keplerian orbital velocity in K(a)dS backgrounds is thus given by the relation
\be                                                                           \label{e14}
{\cal V}^{(\phi)}_{\rm K\pm}(r;\,a,\,y) = \frac{(r^2+a^2)\sqrt{1-yr^3}\mp
           a\sqrt{r}[2+r(r^2+a^2)y]}{\sqrt{\Delta_{r}}[a\sqrt{1-yr^3}\pm r\sqrt{r}]}.
\ee

If the spacetime parameters are appropriately chosen, i.e. the cosmological parameter $y$ enables existence of stable circular orbits and the spin of the K(a)dS black hole $a$ is sufficiently high, the Keplerian orbital velocity on plus-family orbits exhibits non-monotonic, humpy radial profile (``minimum-maximum'' structure) inside an ergosphere of the black hole and close to the marginally stable circular orbit. On the other hand, in KadS ($y<0$) backgrounds the orbital velocity of both families reveals another non-monotonic behaviour outside the ergosphere independently of the spin $a$, caused by an interplay between the cosmic attraction represented by $\Lambda<0$, which strength grows with increasing radius, and an attraction of the central black hole, which strength grows with decreasing radius. Of course, in near-extreme KadS black-hole spacetimes the both types of non-monotonicities in radial profile of the plus-family Keplerian orbital velocity are present; see Fig.~\ref{f1}.

\begin{figure}
\centering
	\includegraphics[width=.6 \hsize]{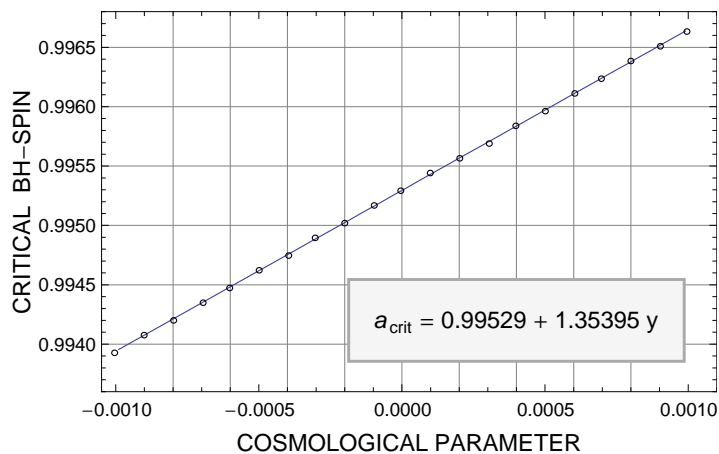}
	\caption{Linear correlation between the cosmological parameter $y$ (cosmological constant $\Lambda$) and the critical spin $a$, for which the ``minimum-maximum'' structure of the Keplerian orbital velocity in K(a)dS backgrounds is getting to exist.}
	\label{f2}
\end{figure}

As M\"{u}ller \& Aschenbach \cite{Mul-Asch:2007:CLAQG:} noted, the linear function fits very well the relation between the critical spin $a_{\rm crit}$, for which the ``minimum-maximum'' structure is getting to exist, and the cosmological parameter $y$ (cosmological constant $\Lambda$) for small positive and negative values of $y$. Taking the exact relation for the Keplerian orbital velocity (\ref{e14}), the corrected linear fit is
\be                                                                           \label{e15}
     a_{\rm crit}=0.99529+1.35395\,y=0.99529+0.45132\,\Lambda.
\ee
It was found by the least-squares method included in \texttt{Wolfram Mathematica 6.0} for numerically computed data-points from the interval of cosmological parameters $-0.001\leq y \leq 0.001$ divided with a step $10^{-4}$; see Fig.~\ref{f2}. However it can be extrapolated satisfactorily up to the range $-0.01\leq y \leq 0.01$, as the relative inaccuracy\footnote{Ratio of the absolute difference between the true and extrapolated value of the critical spin, and the true value.} for $y=0.01$ is $<0.04\,\%$, and for $y=-0.01$ it is $0.1\%$. Note that the range of K(a)dS black-hole parameters, for which the ``minimum-maximum'' structure of the Keplerian orbital velocity exists, is broader than reported in \cite{Mul-Asch:2007:CLAQG:}, and is extended to $a>1$ region, as the highest spin of the  extreme KdS black hole is $a_{\rm BH}\doteq 1.10092$, corresponding to the highest value of the cosmological parameter $y_{\rm BH}\doteq 0.05924$ admitting the KdS black-hole spacetimes \cite{Stu-Sla:2004:PHYSR4:}. We shall report the whole range of K(a)dS spacetime parameters relevant for the humpy profiles of the Keplerian orbital velocity in a subsequent paper being now in preparation. 

\subsection{Non-Keplerian distribution of the specific angular momentum}
The simplest, however unrealistic, non-geodesic circular motion can be characterized by the uniform distribution of the specific angular momentum
\be                                                                           \label{e16}
\ell(r,\,\theta)=\mbox{const}.
\ee
In such a case the orbital velocity in the LNRF is given by the relation
\be                                                                           \label{e17}
{\cal V}^{(\phi)}_{\rm uni}(r,\,\theta;\,a,\,y,\,\ell) = \frac{\rho^2\ell\sqrt{\Delta_{r}\Delta_{\theta}}}{\{A-[\Delta_{\theta}(r^2+a^2)-\Delta_{r}] a\ell\}\sin\theta}.
\ee 

Local extrema of the orbital velocity are located in the equatorial plane ($\theta=\pi/2$), and are given by the condition for the specific angular momentum 
\bea                                                                          \label{e18}
\lefteqn{\ell=\ell_{\rm ex}(r;\,a,\,y)} \\
& & \equiv a+\frac{r^2 [(r^2+a^2)(r-1-yr(2r^2+a^2))-2r\Delta_r]}{ar[r-1-yr(2r^2+a^2)][2+yr(r^2+a^2)] +2a\Delta_r(1-yr^3)}. \nonumber
\eea

\begin{figure}
  \begin{minipage}{.49 \linewidth}
    \centering
    \includegraphics[width=1 \hsize]{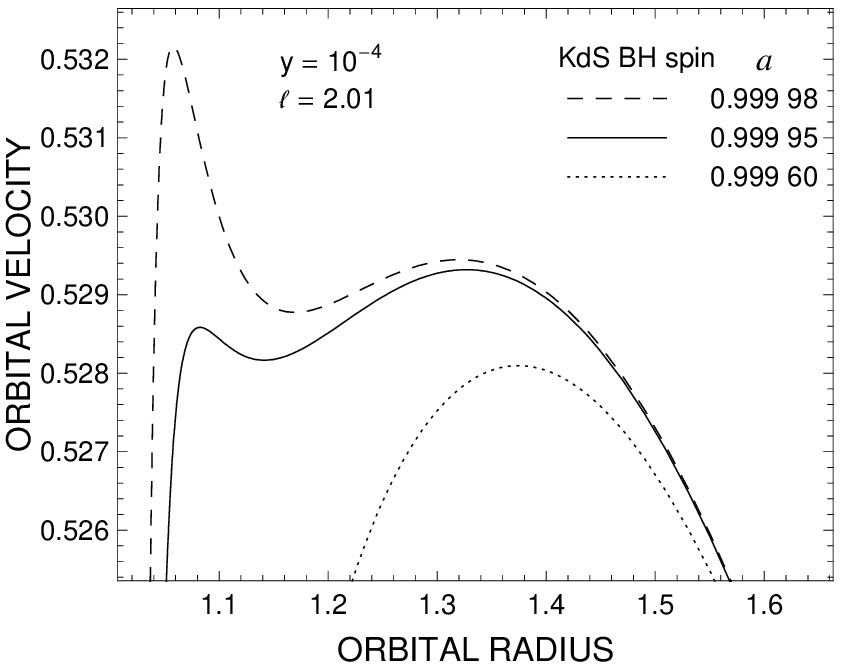}
    \par\small (a)
  \end{minipage}
  \hfill
  \begin{minipage}{.49 \linewidth}
    \centering
    \includegraphics[width=1 \hsize]{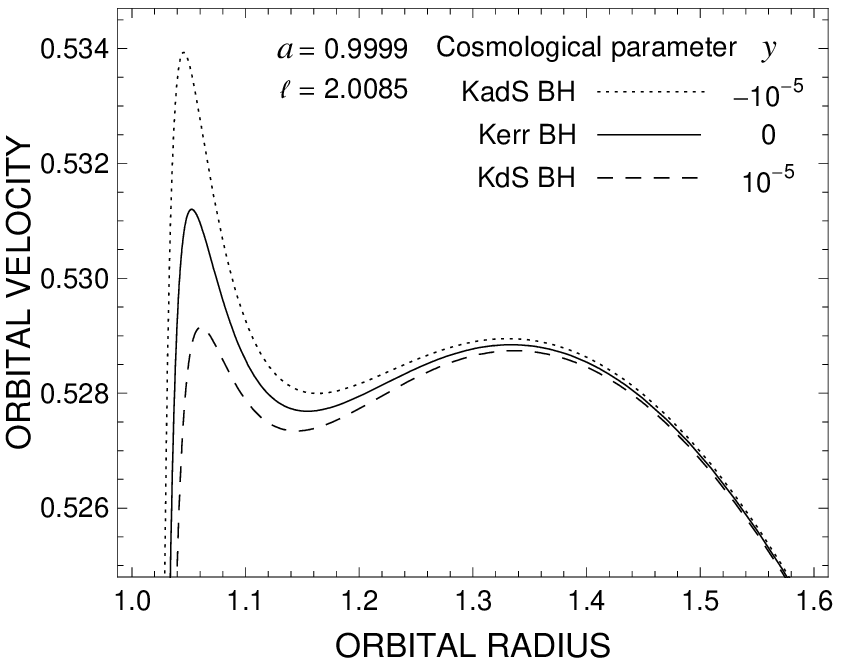}
    \par\small (b)
  \end{minipage}
  \vskip1ex
  \begin{minipage}{.49 \linewidth}
    \centering
    \includegraphics[width=1 \hsize]{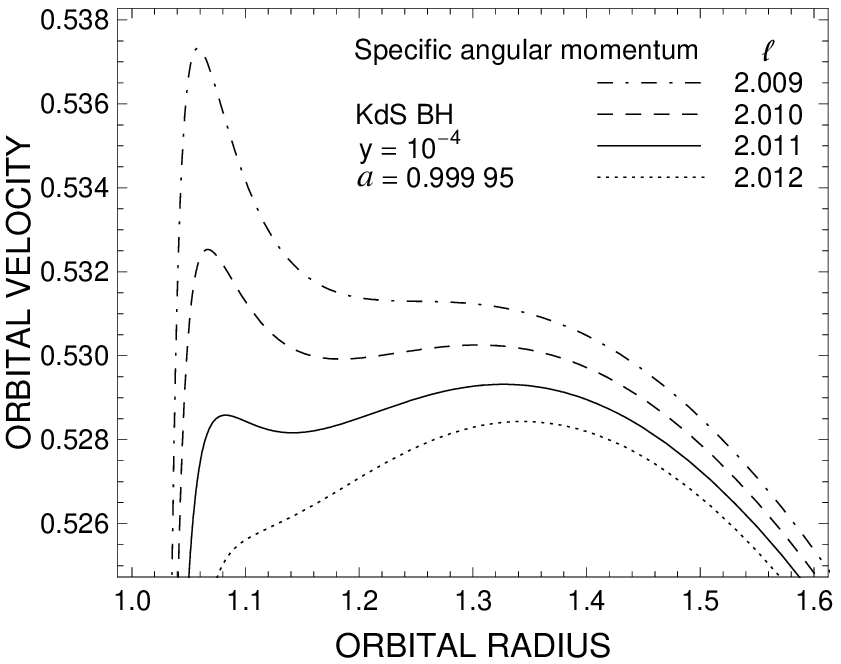}
    \par\small (c)
  \end{minipage}
	\caption{LNRF-related orbital velocity profiles of the $\ell=\mbox{const}$ motion in the equatorial plane of K(a)dS spacetimes. Appropriate combination of spacetime parameters $y,\,a$ together with the specific angular momentum $\ell$ are needed to obtain the humpy profile. The necessary condition, however, is a sufficiently high value of the black-hole spin, $a>a_{\rm crit}$. (a) Equatorial orbital velocity profiles for fixed values of the cosmological parameter $y$ and specific angular momentum $\ell$, and various values of the black-hole spin $a$. (b) Equatorial orbital velocity profiles for fixed values of $a$ and $\ell$, and various values of $y$. (c) Equatorial orbital velocity profiles for fixed values of $a$ and $y$, and various values of $\ell$.}
	\label{f3}
\end{figure}

Radial behaviour of the function $\ell_{\rm ex}(r)$ depends on spacetime parameters $y$ and $a$. In the K(a)dS black-hole spacetimes, in which the humpy orbital-velocity profiles can be found, the $\ell_{\rm ex}$ function is continuous above the black-hole horizon and reveals two positive local extrema: the minimum $\ell_{\rm ex-}$ and the maximum $\ell_{\rm ex+}$. For the specific angular momentum distribution $0<\ell_{\rm ex-}<\ell <\ell_{\rm ex+}$ the orbital velocity profile in the equatorial plane exhibits humpy behaviour, as two local maxima and a local minimum in between them exist. For $0<\ell<\ell_{\rm ex-}$, or $\ell>\ell_{\rm ex+}$, the orbital velocity profile in the equatorial plane exhibits one local maximum only. For $a=a_{\rm crit}$ both local extrema of $\ell_{\rm ex}(r)$ coincide and for $a<a_{\rm crit}$ there is no local extreme in the $\ell_{\rm ex}(r)$ profile. The last two cases correspond to one local extreme in the orbital velocity profile in the equatorial plane: a~maximum (for $\ell>0$), or a~minimum (for $\ell<0$), independently of the spin $a$. Fig.~\ref{f3} shows the spin and $\Lambda$ dependences of the $\ell=\mbox{const}$ orbital velocity profiles in the equatorial plane. 

\begin{figure}
\centering
	\includegraphics[width=.6 \hsize]{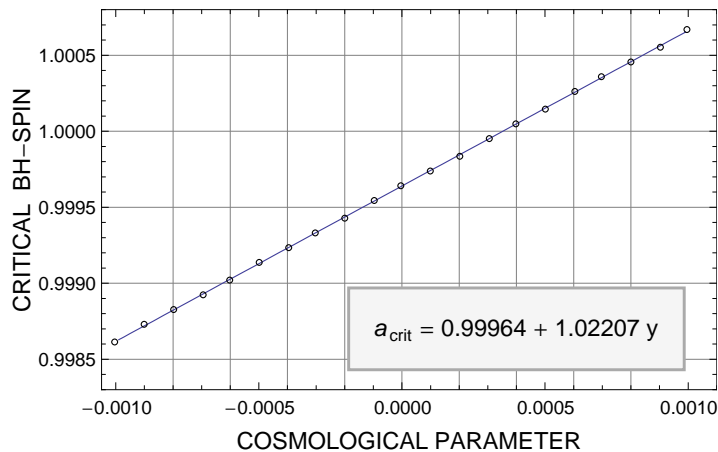}
	\caption{Linear correlation between the cosmological parameter $y$ (cosmological constant $\Lambda$) and the critical spin $a$, for which the ``humpy'' orbital velocity profile of the $\ell=\mbox{const}$ motion in K(a)dS backgrounds is getting to exist.}
	\label{f4}
\end{figure}

For each appropriately chosen, i.e. sufficiently small in its absolute value, cosmological parameter $y$, there again exists the minimal value of the spin, $a_{\rm crit}$, necessary for existence of the humpy orbital-velocity profiles of the $\ell=\mbox{const}$ motion. As in the case of Keplerian motion, a~linear correlation between the critical spin and the cosmological parameter can be found, being described by the relation
\be                                                                           \label{e19}
     a_{\rm crit}=0.99964+1.02207\,y=0.99964+0.34069\,\Lambda.
\ee  
The linear fit was found by the same method as the previous one for the Keplerian motion, and for the same range of cosmological parameters $-0.001\leq y \leq 0.001$; see Fig.~\ref{f4}. However, it can be again extrapolated satisfactorily up to the range $-0.01\leq y \leq 0.01$; the relative inaccuracy for $y=\pm 0.01$ is $\sim 0.05\,\%$. Presented linear relation completely differs from that given in \cite{Mul-Asch:2007:CLAQG:}. Moreover the range of spacetime parameters, for which the emerged humpy behaviour exists, is sufficiently broader than that reported in \cite{Mul-Asch:2007:CLAQG:}. The whole range including the naked singularity cases will be reported in the paper in preparation. 

\section{Conclusions}
Non-monotonic (humpy) profiles of the LNRF-related orbital velocity for the Keplerian and $\ell=\mbox{const}$ motion in K(a)dS spacetimes are analyzed, and the results are confronted with those of M\"{u}ller \& Aschenbach \cite{Mul-Asch:2007:CLAQG:}. 

At first it should be noted that M\"{u}ller \& Aschenbach used wrong restrictive condition $|a|\leq 1$ for the existence of K(a)dS black holes. In KdS ($\Lambda>0$) spacetimes the limiting spin corresponding to the extreme black hole follows a condition $1<a_{\rm BH}(\Lambda)<1.10092$, see \cite{Stu-Sla:2004:PHYSR4:}, while in KadS ($\Lambda<0$) spacetimes $a_{\rm BH}(\Lambda)<1$, and depends on the value of $\Lambda$. Omitting of this fact influences substantially the range of spacetime parameters $a,\,\Lambda$ for which the ``minimum-maximum'' structure in the orbital velocity profiles exists.

Linear fits between the black-hole spin and the cosmological constant for the onset of humpy structure in radial profiles of the LNRF-related orbital velocity obtained here and in \cite{Mul-Asch:2007:CLAQG:} differ substantially. In the case of Keplerian angular momentum distribution the difference appears in the slope, and this is caused clearly by the fact that M\"{u}ller \& Aschenbach used an incorrect formula for the Keplerian angular velocity in K(a)dS spacetimes, see e.g. \cite{Sla-Stu:2005:CLAQG:} for correct formula. In the case of uniform angular momentum distribution ($\ell=\mbox{const}$) the situation is more complex and direct comparison of the results presented here and in \cite{Mul-Asch:2007:CLAQG:} cannot be done, since M\"{u}ller \& Aschenbach do not give full information about conditions employed in construction of their fit. In our result, both the critical value $a_{\rm crit}(\Lambda=0)$ and the slope are different than those in \cite{Mul-Asch:2007:CLAQG:}. Critical spins presented here for matter configurations with $\ell=\mbox{const}$ angular momentum distribution are determined without invoking any other physical conditions on the configurations. 

Our analysis leads to the following conclusions. We show that the ranges of spacetime parameters, for which the humpy orbital velocity profiles  of Keplerian and $\ell=\mbox{const}$ motion exist, are broader than reported in \cite{Mul-Asch:2007:CLAQG:}. 
Comparing the orbital velocity profiles in the Kerr and K(a)dS backgrounds we conclude that in the case of Keplerian motion the repulsive cosmological constant, $\Lambda >0$, decreases the depth of the hump, while the attractive cosmological constant, $\Lambda <0$, increases the velocity difference in the hump.\footnote{The same result is also presented by M\"{u}ller \& Aschenbach \cite{Mul-Asch:2007:CLAQG:}.} On the other hand, in the case of $\ell=\mbox{const}$ motion we obtained inverse behaviour of the humpy orbital velocity profiles in dependence of the cosmological constant, see Fig.~\ref{f3}, contrary of the work \cite{Mul-Asch:2007:CLAQG:}. For both analyzed types of motion the decreasing of the cosmological constant $\Lambda$ causes similar qualitative changes in the orbital velocity profiles like an increasing of the BH-spin $a$. This result is in accord with the linear correlation between the critical spin and the cosmological parameter for both Keplerian and $\ell=\mbox{const}$ motions (Figs.~\ref{f2},~\ref{f4}). Moreover in the case of $\ell=\mbox{const}$ motion the analogical qualitative behaviour of the orbital velocity profiles takes place when the specific angular momentum $\ell$ decreases for given values of $\Lambda$ and $a$. Further we would like to stress that due to the attractive cosmological constant, $\Lambda <0$, there is a non-monotonic orbital velocity profile of the Keplerian motion in all KadS spacetimes allowing stable circular orbits, independently of the spin $a$. 

Although the analysis is done only for K(a)dS black holes, it can be extended also to naked singularity backgrounds. As the main aim of this paper is to briefly comment and correct the recently published work of M\"{u}ller \& Aschenbach \cite{Mul-Asch:2007:CLAQG:}, we postpone the analysis of the LNRF-related orbital velocity profiles in K(a)dS naked-singularity backgrounds to the subsequent paper. Moreover, in the Kerr background it was suggested by Aschenbach \cite{Asch:2004:ASTRA:}, and later by Stuchl\'{\i}k et al. \cite{Stu-Sla-Tor:2007a:ASTRA:,Stu-Sla-Tor:2007b:ASTRA:} that the orbital velocity hump could be related with processes capable to excite the epicyclic oscillations in accretion-disc systems around near-extreme black holes, and in this sence connected with quasi-periodic oscillations observed in some near-extreme Kerr black hole candidates. Influence of the cosmological constant on the so-called ``humpy frequency'', as well as the relevance of the humpy effect for $\ell=\mbox{const}$ barotropic perfect fluid tori in K(a)dS backgrounds is also in preparation.

\section*{Acknowledgment}
The authors greatfully acknowledge financial support from the Czech Government (MSM~4781305903).

\bibliographystyle{unsrt} 
\section*{References}

\end{document}